\documentclass[pra,superscriptaddress,twocolumn,amsmath,amssymb]{revtex4-1} 
\usepackage{graphicx}
\usepackage{dcolumn}
\usepackage{bm}
\usepackage{amsmath}
\usepackage{float}
\usepackage{mathrsfs}

\begin{document}

\title{Spectrally resolved Hong-Ou-Mandel interference between independent sources}

\author{Rui-Bo Jin}
\email{ruibo@nict.go.jp}
\affiliation{National Institute of Information and Communications Technology (NICT), 4-2-1 Nukui-Kitamachi, Koganei, Tokyo 184-8795, Japan}
\author{Thomas Gerrits}
\affiliation{National Institute of Standards and Technology (NIST), 325 Broadway, Boulder, Colorado 80305, USA}
\author{Mikio Fujiwara}
\affiliation{National Institute of Information and Communications Technology (NICT), 4-2-1 Nukui-Kitamachi, Koganei, Tokyo 184-8795, Japan}
\author{Ryota Wakabayashi}
\affiliation{National Institute of Information and Communications Technology (NICT), 4-2-1 Nukui-Kitamachi, Koganei, Tokyo 184-8795, Japan}
\affiliation{Waseda University, 3-4-1 Okubo, Shinjyuku, Tokyo 165-8555, Japan}
\author{Taro Yamashita}
\affiliation{National Institute of Information and Communications Technology (NICT), 588-2 Iwaoka, Kobe 651-2492, Japan}
\author{Shigehito Miki}
\affiliation{National Institute of Information and Communications Technology (NICT), 588-2 Iwaoka, Kobe 651-2492, Japan}
\author{Hirotaka Terai}
\affiliation{National Institute of Information and Communications Technology (NICT), 588-2 Iwaoka, Kobe 651-2492, Japan}
\author{Ryosuke Shimizu}
\affiliation{ University of Electro-Communications (UEC), 1-5-1 Chofugaoka, Chofu,  Tokyo 182-8585, Japan}
\author{Masahiro Takeoka}
\affiliation{National Institute of Information and Communications Technology (NICT), 4-2-1 Nukui-Kitamachi, Koganei, Tokyo 184-8795, Japan}
\author{Masahide Sasaki}
\affiliation{National Institute of Information and Communications Technology (NICT), 4-2-1 Nukui-Kitamachi, Koganei, Tokyo 184-8795, Japan}

\date{\today }

\begin{abstract}
Hong-Ou-Mandel (HOM) interference  between independent photon sources (HOMI-IPS) is the fundamental block for quantum information processing, such as quantum gate, Shor's algorithm, Boson sampling, etc. All the previous HOMI-IPS experiments were carried out in time-domain, however, the spectral information during the interference was lost, due to technical difficulties. Here, we investigate the HOMI-IPS in spectral domain using the recently developed fast fiber spectrometer, and demonstrate the spectral distribution during the HOM interference between two heralded single-photon sources, and two thermal sources. This experiment can not only deepen our understanding of HOMI-IPS in the spectral domain, but also be utilized to improve the visibility by post-processing spectral filtering.
\end{abstract}

\maketitle

\section{Introduction}
Hong-Ou-Mandel(HOM) interference \cite{Hong1987} between independent photon sources (HOMI-IPS) is at the heart of quantum information processing involving the quantum interference of single photons \cite{Walmsley2005, Lu2007a}.
Scaling such systems to large dimensions requires many single-photons to interfere in a single mode \cite{Broome2013, Spring2013}.
This establishes a fundamental challenge since the outcome fidelity of the quantum processor depends on the interference visibility of all photons in all modes.
To insure high output-fidelity the interference of at least two independent photons must approach unity.

A variety of  HOM-IPS experiments have been demonstrated at near-infrared wavelengths \cite{Kaltenbaek2006, Mosley2008a, Mosley2008b, Rarity2005, Jin2011, Soller2011, Tanida2012, Zhao2014} or telecom wavelengths \cite{Takesue2007, Xue2010, Aboussouan2010, Harada2011, Jin2013PRA, Harder2013, Bruno2014}, and with heralded single-photon sources \cite{Kaltenbaek2006, Mosley2008a, Mosley2008b, Jin2013PRA, Harder2013, Bruno2014}, weak coherent sources \cite{Rarity2005, Jin2011}, or even thermal sources \cite{Li2008, Jin2013PRA}.
In all the previous HOMI-IPS experiments, the HOM dips are obtained by recording the  coincidence counts as a function of the temporal delay.
The knowledge of the spectral correlations of the interfering photons can reveal some important information about the interference visibility \cite{MosleyPhD}.
In the past, the spectral correlation content could not be observed due to the long acquisition times required when measuring spectral correlations using two tunable bandpass filters.

Recently, a new technique for measuring spectral correlations in a HOM interference was presented by Gerrits, et al \cite{Gerrits2015, Shalm2013}.
This technique  has the merits of high speed and high spectral resolution, and therefore allows measuring and analyzing the spectral correlations from a HOM interference. 

Here, we measure the spectral correlation between the two output ports of the HOMI-IPS beam splitter with two different photon sources (heralded single-photon sources, and thermal sources) using the method developed in \cite{Gerrits2015}.
This experiment can help deepen our understanding of HOMI-IPS from the viewpoint of spectral domain and allows improving the HOM-IPS visibility  by post-processing spectral filtering.

First, we present the theory and simulation for HOM interference between two heralded single-photon states, and between two thermal states. We then present our experimental results for these two kinds of HOMI-IPS experiments. The appendix of this manuscript contains our theoretical model for HOM interference between two heralded single-photon states, by considering the infinitely higher order components in the photon sources.

\section{Theory and simulation}

\subsection{Theory of four-fold HOM interference between two heralded single-photon states}
Two independent spontaneous parametric down conversion (SPDC) sources are used in the experiment.
The signal and idler photons from the first and the second SPDC source have the joint spectral amplitude of $f_1 (\omega _{s1} ,\omega _{i1})$ and $f_2 (\omega _{s2} ,\omega _{i2})$, where $\omega$ is the angular frequency.  The subscripts $1$ and $2$ represent the first and the second SPDC, respectively;  $s$ and $i$ denote the down-converted photons, i.e., the signal and idler.
The signal photons ($s1$ and $s2$) are sent to a 50/50 beamsplitter for HOM interference, while the idler photons ($i1$ and $i2$) function as heralders, i.e., detected by single-photon detectors to claim the existence of $s1$ and $s2$ (see the layout in Fig. \ref{setup}).
The four-fold coincidence probability $P_4 (\tau )$ as a function of the time delay $\tau $ can be written as \cite{MosleyPhD}:
\begin{equation}\label{eq:P4}
   P_4 (\tau ) = \frac{1}{4}\int\limits_0^\infty  {\int\limits_0^\infty  {\int\limits_0^\infty  {\int\limits_0^\infty  {d\omega _{s1} d\omega _{s2} d\omega _{i1} d\omega _{i2} } } } }          I_4 ( \tau ),
\end{equation}
where,
\begin{equation}\label{eq:I4}
\begin{array}{ll}
   I_4 (\tau ) = & {\rm{|}}f_1 (\omega _{s1} ,\omega _{i1} )f_2 (\omega _{s2} ,\omega _{i2} ) - \\
                & f_1 (\omega _{s2} ,\omega _{i1} )f_2 (\omega _{s1} ,\omega _{i2} )e^{ - i(\omega _{s2}  - \omega _{s1} )\tau } {\rm{|}}^{\rm{2}}.  \\
\end{array}
\end{equation}
The correlated spectral intensity (CSI),   $\tilde I _4 (\tau )$, for the two beam splitter output ports  at time delay $\tau$ can be expressed as:
\begin{equation}\label{eq:I4-2}
   \tilde I _4 (\tau ) = \int\limits_0^\infty  {\int\limits_0^\infty    { d\omega _{i1} d\omega _{i2} } }   I_4 ( \tau ).
\end{equation}

\begin{figure*}[htbp]
\centering
\includegraphics[width=12.5 cm]{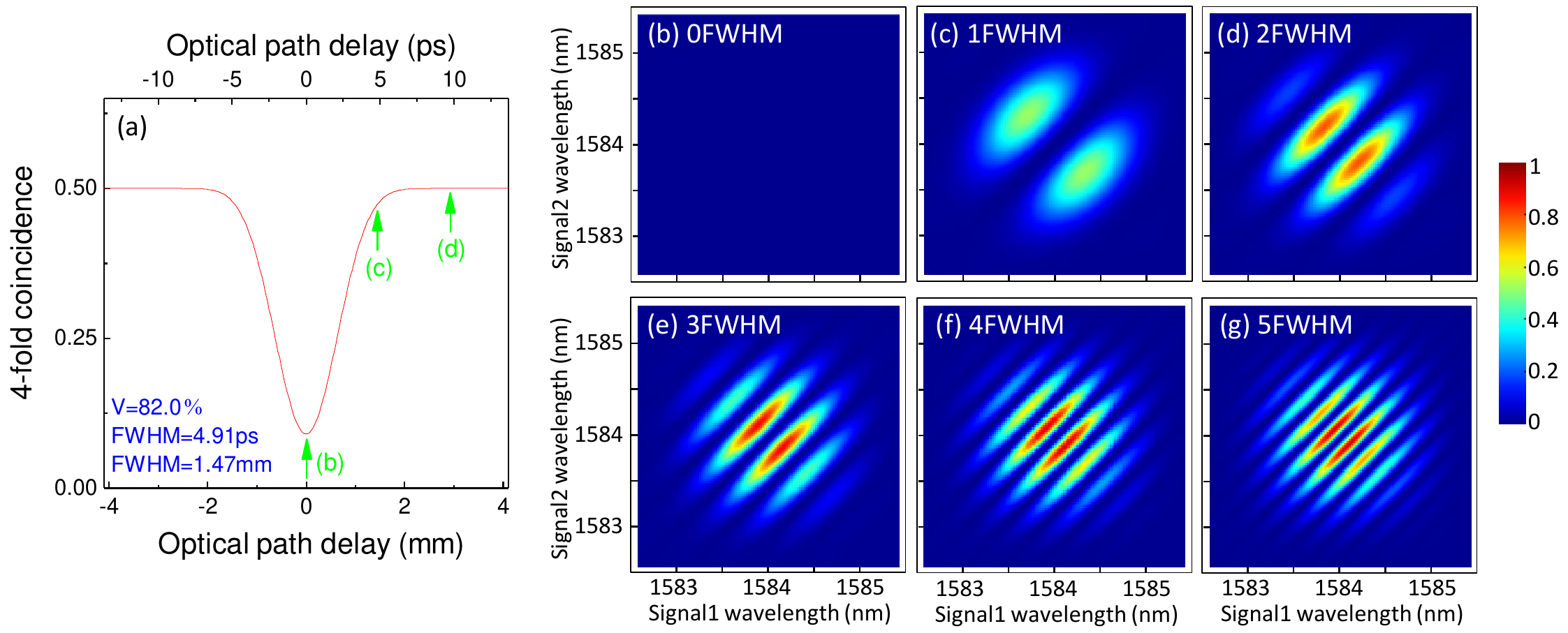}
\caption{ The simulation for the HOM interference between two heralded single-photon states.(a) The simulated HOM dip. (b-e) Simulated correlated spectral intensity (CSI) at different delay positions in the HOM dip.
  } \label{simu1}
\end{figure*}

Figure \ref{simu1} shows the simulated HOM dip and CSI using the experimental parameters of the setup: $\sim$ 2 ps pump pulses at 792 nm, 30-mm-long   periodically poled $\mathrm{KTiOPO_4}$ (PPKTP) crystal.
As shown in Fig. \ref{simu1}(a), without any bandpass filtering, the visibility of the four-fold HOM dip is 82.0\%, and the FWHM (full width at half maximum) is 4.91 ps (1.47 mm).
The intrinsic  spectral purity of this source is 0.82 \cite{Jin2013OE, Jin2014OE}, which is also the upper bound of the four-fold HOM  interference visibility in our case \cite{Mosley2008a, Jin2013PRA}.
The theoretical prediction in Fig. \ref{simu1}(b-g) shows fringes in the CSI at different delay positions. The fringe frequency increases with increasing distance from the center of the HOM dip.

\subsection{Theory of two-fold HOM interference between two thermal states}
The two-fold coincidence probability $ P_{t}(\tau )$ between signal1 and signal2 (without the gating of idler1 and idler2, signal1 and signal2 are thermal states) as a function of the time delay $\tau$ can be written as \cite{Ou2007}
\begin{equation}\label{eq:A1}
P_{t} (\tau ) = \frac{1}{4}\int\limits_0^\infty  {\int\limits_0^\infty  {\int\limits_0^\infty  {\int\limits_0^\infty  {d\omega _{s} d\omega _{s^,} d\omega _{i} d\omega _{i^,} } } } } I_{t} (\tau ),
\end{equation}
where,
\begin{equation}\label{eq:A2}
I_{t} (\tau ) =\mathscr{A1} +\mathscr{E} + \mathscr{A2} - \mathscr{E2}(\tau ),
\end{equation}
with
\begin{equation}\label{eq:A3}
\mathscr{A}=2\left| {f (\omega _{s} ,\omega _{i} )f (\omega _{s^,} ,\omega _{i^,} )} \right|^2,
\end{equation}
\begin{equation}\label{eq:A4}
\mathscr{E}=2f (\omega _{s} ,\omega _{i} )f (\omega _{s^,} ,\omega _{i^,} )f^*(\omega _{s} ,\omega _{i^,} )f^*(\omega _{s^,} ,\omega _{i} ),
\end{equation}
\begin{equation}\label{eq:A5}
\begin{array}{l}
\mathscr{E}(\tau )= \\
2 f (\omega _{s} ,\omega _{i} )f (\omega _{s^,} ,\omega _{i^,} )f^*(\omega _{s} ,\omega _{i^,} )f^*(\omega _{s^,} ,\omega _{i} )  e^{-i(\omega _{s}  - \omega _{s^,}) \tau },
\end{array}
\end{equation}
where   $\mathscr{E}(0)=\mathscr{E}$, $\mathscr{E}(\infty)=0$,  and  $\mathscr{E}  \le \mathscr{A}$. 
For simplicity, here we assume two nonlinear crystals are identical:  $f_1(\omega _{s1} ,\omega _{i1} )=f_2 (\omega _{s2} ,\omega _{i2} )=f (\omega _{s} ,\omega _{i} )$.
$f^*(\omega _{s} ,\omega _{i} )$ is the complex conjugate of $f(\omega _{s} ,\omega _{i} )$.
In Eq.(\ref{eq:A2}),  $\mathscr{A} - \mathscr{E}(\tau )$ is the interference parts, which is contributed by single photons from two different crystals,
and $\mathscr{A} +\mathscr{E}$ is the background parts, which is contributed by two photons from one crystal.

In fact, $I_4 (\tau )$ in Eq.(\ref{eq:I4}) can be expressed as \cite{Ou2007}
\begin{equation}\label{eq:A6}
 I_4 (\tau ) =\mathscr{A} - \mathscr{E}(\tau ).
\end{equation}
Eq.(\ref{eq:A2}) and Eq.(\ref{eq:A6}) suggest  the HOM interference between two thermal states is including the HOM interference between two heralded single-photon states $\mathscr{A} - \mathscr{E}(\tau )$, plus a constant background term of $\mathscr{A} +\mathscr{E}$.
The correlated spectral intensity (CSI),   $\tilde I _t (\tau )$, for the two beam splitter output ports  at time delay $\tau$ can be expressed as:
\begin{equation}\label{eq:A2-2}
   \tilde I _t (\tau ) = \int\limits_0^\infty  {\int\limits_0^\infty    { d\omega _{i} d\omega _{i^,} } }   I_t ( \tau ).
\end{equation}

\begin{figure*}[tbp]
\centering\includegraphics[width=12.5 cm]{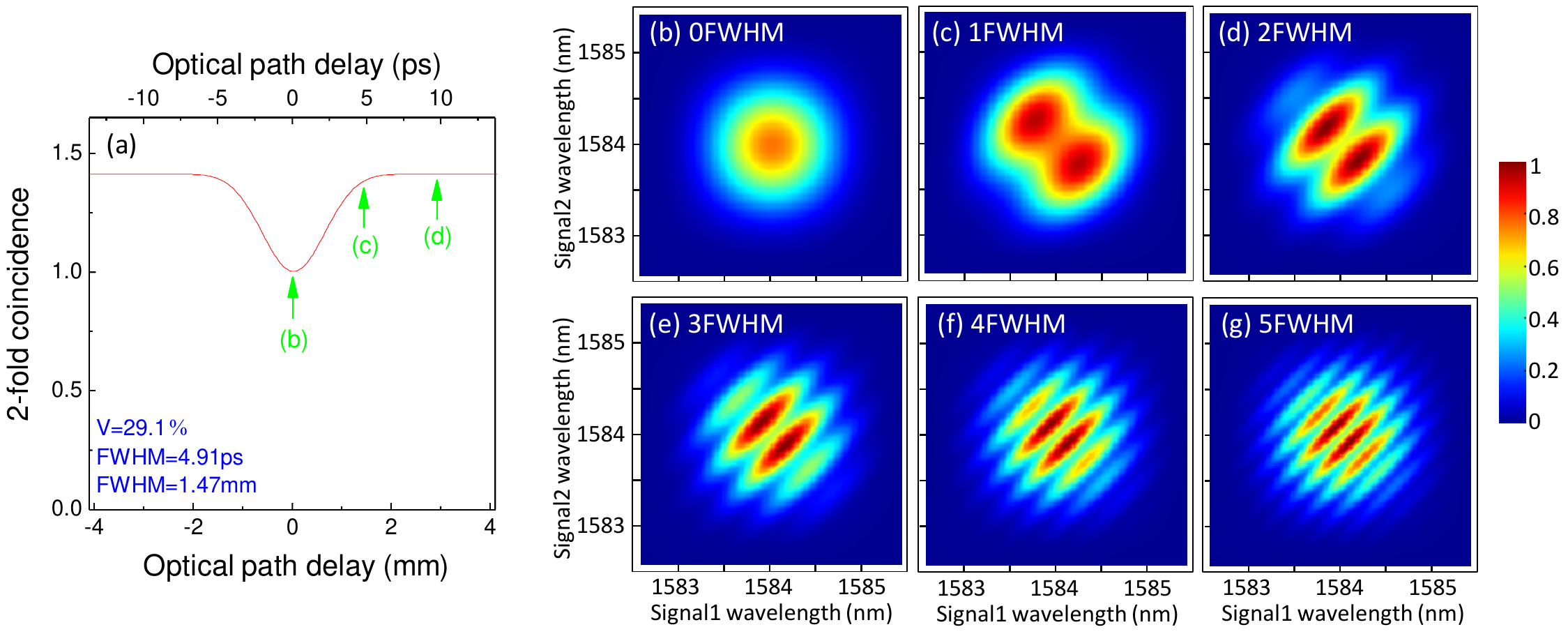}
\caption{ The simulation for the HOM interference between two thermal states.  (a) The simulated two-fold HOM dip.(b-g) CSI at different delay positions in the HOM dip.
  } \label{simu2}
\end{figure*}

With the same condition for Fig. \ref{simu1}, we simulate the two-fold HOM dip in Fig. \ref{simu2}(a), and $ \tilde  I_{t} (\tau )$   at different delay positions are shown in Fig. \ref{simu2}(b-g).
The visibility $V$ of HOM interference between two thermal state has an upper bound of $\frac{1}{3}$ \cite{Ou2007}:
\begin{equation}\label{eq:A7}
V = \frac{{P_t (\infty ) - P_t (0)}}{{P_t (\infty )}} = \frac{\mathscr{E}}{{2\mathscr{A} + \mathscr{E}}} \le \frac{1}{3},
\end{equation}
where $P_t (\pm \infty )  \propto \frac{1}{4}(2\mathscr{A} + \mathscr{E})$ and  $P_t (0 )  \propto \frac{1}{4}(2\mathscr{A})$.
Refer to \cite{Ou2007} for more details of the theory of interference between two thermal states.

\section{Experiment and results}
\subsection{Experiment of four-fold HOM interference between two heralded single-photon states}
\begin{figure*}[tbp]
\centering\includegraphics[width=12.5 cm]{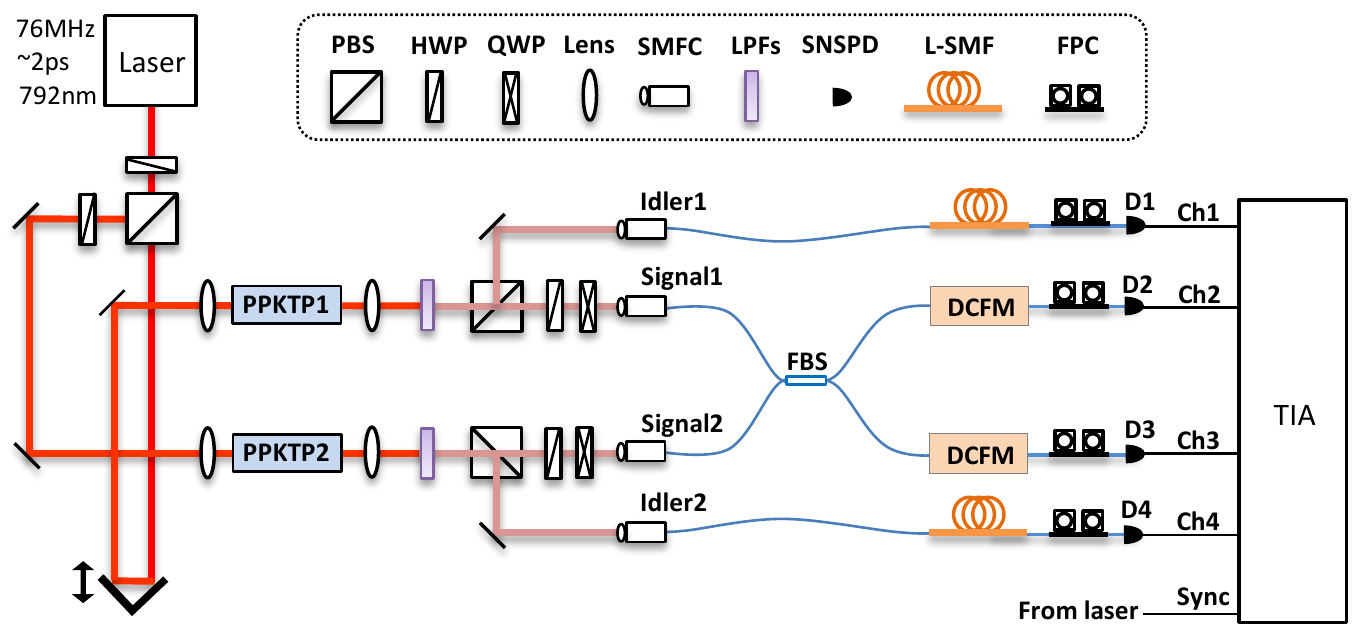}
\caption{ The experimental setup. Picosecond laser pulses (76 MHz, 792 nm, temporal duration $\sim$ 2 ps) from a mode-locked Titanium sapphire laser (Mira900, Coherent Inc.) were divided by a polarizing beam splitter (PBS) into two paths, and pumped two 30-mm-long PPKTP crystals with a poling period of 46.1 $\mu$m for type-II SPDC.
The downconverted signal (horizontal polarization) and idler  (vertical polarization) photons with a degenerate wavelength of 1584 nm were divided by two PBSs, and then coupled into single-mode fibers (SMFC).
The signal photons were sent to a 50/50 fiber beam splitter (FBS) and connected to two dispersion compensation fiber modules (DCFM), 
in order to generate large dispersion. The idler photons were connected to two long-distance single-mode fiber spools (L-SMF), 
to achieve the same path delay as the signal photons.
Finally, all collected photons were sent to four superconducting nanowire single-photon detectors (SNSPDs), which were connected to four input channels of a time interval analyzer (TIA).
Our SNSPDs have a system detection efficiency  of around 70\% with a dark count rate less than 1 kcps \cite{Miki2013, Yamashita2013, Jin2015OC}.
Electrical signals recording the timing information of the pump pulses were sent to the TIA for synchronization (Sync).
LPF = long pass filters, HWP = half-wave plate, QWP = quarter-wave plate, FPC = fiber polarization controller.
   } \label{setup}
\end{figure*}

The experimental setup is shown in Fig. \ref{setup}.
The experimental setup is the extension of our previous work \cite{Jin2013PRA}.
We add two kinds of dispersion fibers to the four detection channels.
Each output port of the fiber beam splitter (FBS) is connected to a dispersion compensation fiber module (DCFM),
a 7.53-km-long single-mode fiber (SMF) with high dispersion designed to compensate the dispersion of a 50-km-long commercial SMF at the L-band of telecom wavelengths.
The DCFM has a dispersion of 125.0 ps/km/nm and an insertion loss of 5.2 dB.
Due to the large dispersion, the DCFM converts the spectral distribution information of the photons into arrival time information, which can be determined with high accuracy using superconducting nanowire single-photon detectors (SNSPDs).
The DCFMs, SNSPDs and the  time interval analyzer (TIA) constitute a fiber spectrometer\cite{Avenhaus2009, Gerrits2011}.
To ensure that the idler photons have the same optical path delay as the signal photons, we add two 7.53-km-long commercial SMFs (L-SMF in Fig. \ref{setup}) to the idler channels.
L-SMF has a dispersion of 27.3 ps/km/nm and an insertion loss of 1.6 dB.
Assuming a system jitter of 100 ps, the resolution of our fiber spectrometer is approximately $0.11$ nm.

\begin{figure*}[tbp]
\centering\includegraphics[width=12.5 cm]{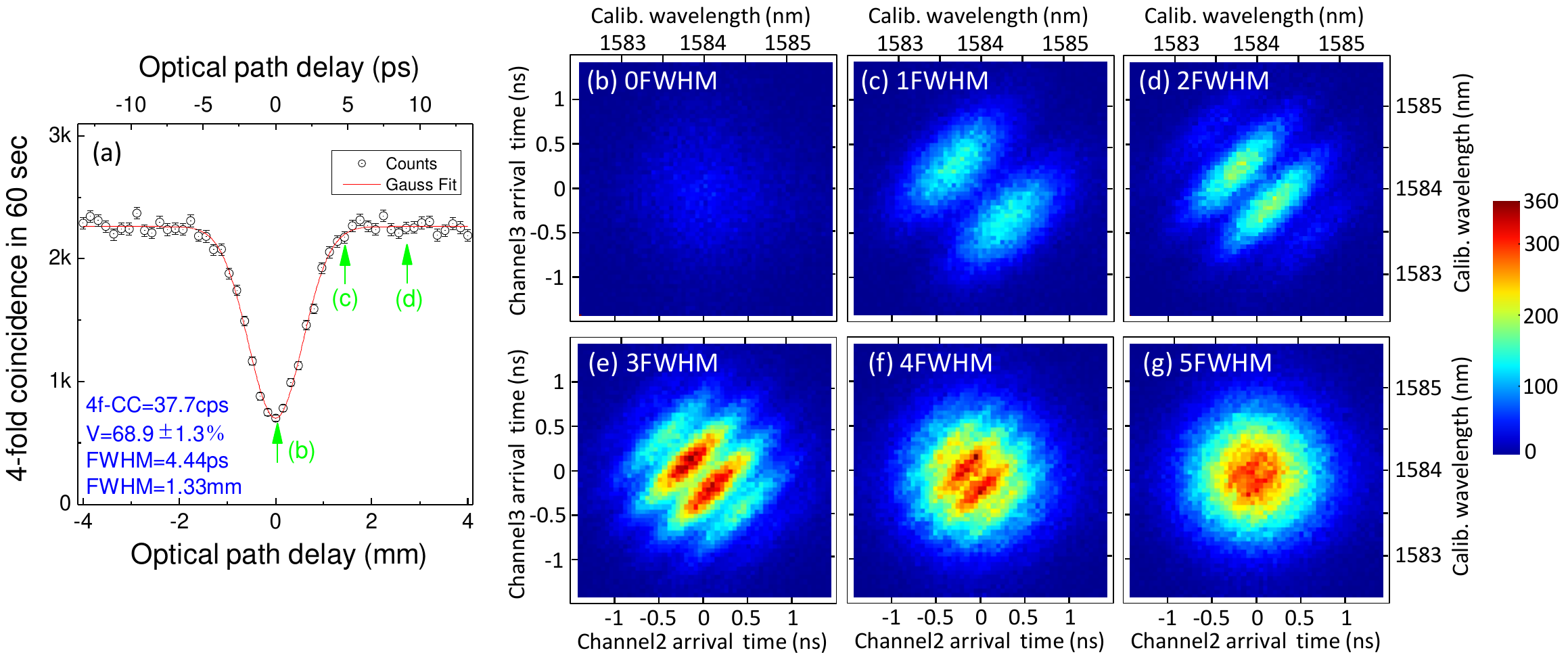}
\caption{ (a) The measured four-fold HOM dip between signal1 and signal2, with idler1 and idler2 as herladers. (b-e) Measured correlated spectral intensity (CSI) at different delay positions in the HOM dip. Each CSI data was accumulated in 2 hours. 
  } \label{result1}
\end{figure*}

We pump the PPKTP crystal sources with a Ti:Sapphire oscillator at $76$ MHz, each with an average pump power of 100 mW. We obtain a singles count rate of  $\sim$ 1.6 Mcps, a two-fold coincidence count rate (2f-CC) of 350 kcps, and a four-fold coincidence count rate (4f-CC) of 880 cps. To our knowledge, this 4f-CC is the highest rate reported at telecom wavelengths.
The mean photon number per pulse and the total efficiency of
each spatial mode are then estimated to be $\bar{n}=0.114$ and $\eta=0.19$,
respectively.
After insertion of the  DCFM and the L-SMF, the 2f-CC and 4f-CC are reduced to 35 kcps and 37.7 cps.
First, we measure a HOM interference between signal1 and signal2, with idler1 and idler2 as heralders. The result is shown in Fig. \ref{result1}(a).
Without subtraction of any background counts, the raw visibility of the HOM dip in Fig. \ref{result1}(a) is 68.9$\pm$1.3\%, which is consistent with our previous results in  \cite{Jin2013PRA, Jin2015SR}.
The degradation from the expected visibility of 82.0\% (Fig. \ref{simu1}(a)) to 68.9$\pm$1.3\% is mainly caused by the multi-pair emission in the SPDC \cite{Takeoka2015}.
This is confirmed by the numerical simulation.
Based on the theory in \cite{Takeoka2015}, we develop a theoretical model
including losses, multi-pair photon emissions (up to infinitely higher
order photons), and the realistic frequency mode structure, i.e.
the joint spectral amplitude (see Appendix).
For $\bar{n}=0.114$ and $\eta=0.19$, the visibility is simulated to be
68.1\% which agrees with the experimental result.
Note that in principle, the visibility can be further improved by decreasing the pump power, but with the tradeoff of lower count rate.
The FWHM of this HOM dip is 4.44 ps (1.33 mm), which corresponds well with the theoretical prediction in Fig. \ref{simu1}(a).

Figure \ref{result1}(b-g) shows the measurements of the CSI at different delay positions, 0 FWHM to 5 FWHMs away from the dip center.
The fringe observed at the HOM dip is very weak (shown in Fig. \ref{result1}(b)), as expected from Eq.(\ref{eq:I4-2}).
The CSI clearly shows fringes at delay positions from 1FWHM to 4FWHM in Fig. \ref{result1}(c-f).
The shape of the CSI is almost circular at 5FWHM (Fig. \ref{result1}(g)), and fringes are not observed due to the limited resolution of our fiber spectrometer. %
The observed CSI from 0FWHM to 4FWHM  agree well with our theoretical predictions in Fig. \ref{simu1}(b-f).

\subsection{Experiment of two-fold HOM interference between two thermal states}
\begin{figure*}[tbp]
\centering\includegraphics[width=12.5 cm]{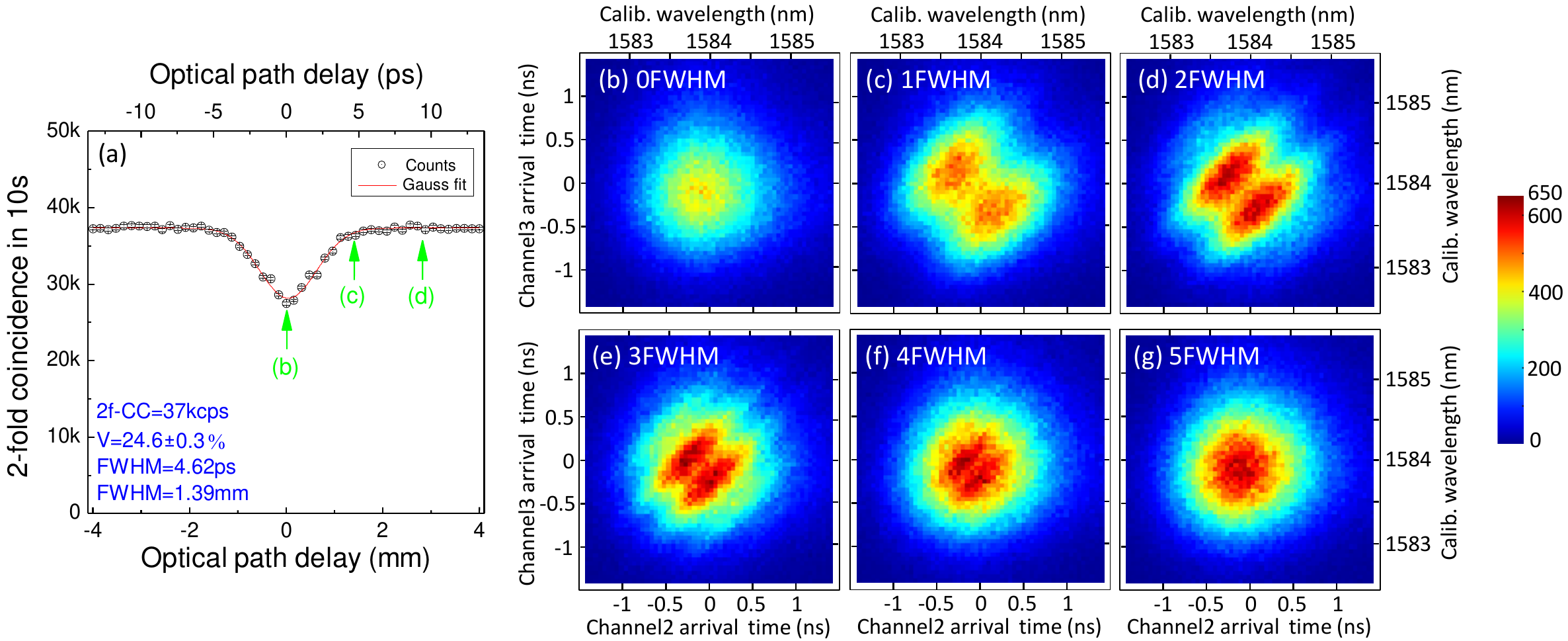}
\caption{ (a) The measured two-fold HOM dip between signal1 and signal2. (b-e) Measured CSI at different delay positions in the HOM dip. Each data was accumulated in  10 minutes.
  } \label{result2}
\end{figure*}

In the following we present our results when measuring the two-fold coincidences between signal1 and signal2, without the heralding of idler1 and idler2, i.e., two thermal states.
The measured HOM dip is shown in Fig. \ref{result2}(a), with a visibility of 24.6$\pm$0.3\% and FWHM of 4.62 ps (1.39 mm).
This obtained visibility is comparable to our previous results in \cite{Jin2013PRA}.
The measured CSI data are shown in  Fig. \ref{result2}(b-g), with accumulation time of 10 minutes for each data set.
Surprisingly, the CSIs also show fringes at the different delay positions from 1FWHM to 4FWHM, even with an interference visibility is as low as 24.6$\pm$0.3\%.
These measured results  correspond well with our simulated results in Fig. \ref{simu2}.
The degradation from the expected visibility of 29.1\% (Fig. \ref{simu2}(a)) to 24.6$\pm$0.3\% is mainly caused by the limited indistinguishability of the two different SPDC sources.

\section{Discussion and Outlook}
It is meaningful to compare the HOMI-IPS (using two SPDCs, e.g. the case in this experiment) to the HOM interference using a twin photon source (using only one SPDC, e.g. the case in \cite{Hong1987}).
In the latter case, the two-fold coincidence probability $ P_2(\tau )$ as a function of the time delay $\tau$ in the HOM interference is given by \cite{Grice1997, Wang2006, JinPhD}:
\begin{equation}\label{eq:P2}
P_2(\tau ) =   \frac{1}{4} \int\limits_0^\infty  \int\limits_0^\infty  d\omega _s  d\omega _i \left| {[f(\omega _s ,\omega _i ) - f(\omega _i ,\omega _s )e^{ - i(\omega _s  - \omega _i )\tau } ]} \right|^2.
\end{equation}
The CSI of the two beam splitter output ports $I(\tau )$ in this HOM interference can be expressed as:
\begin{equation}\label{eq:I2}
\begin{array}{l}
I_2(\tau ) =  \left| {[f(\omega _s ,\omega _i ) - f(\omega _i ,\omega _s )e^{ - i(\omega _s  - \omega _i )\tau } ]} \right|^2.
\end{array}
\end{equation}
To achieve perfect HOM interference, Eq.(\ref{eq:P2}) requires symmetry of the joint spectral distribution, i.e., $f(\omega _s ,\omega _i ) = f(\omega _i ,\omega _s )$, but doesn't require high spectral purity.
However,  to achieve high visibility, Eq.(\ref{eq:P4}) requires not only high symmetry,  but also a high spectral purity.
In Eq.(\ref{eq:P4}), to obtain 100\% visibility, i.e., $P_4(0) = 0$,  the condition of $f_1 (\omega _{s1} ,\omega _{i1} )f_2 (\omega _{s2} ,\omega _{i2} ) = f_1 (\omega _{s2} ,\omega _{i1} )f_2 (\omega _{s1} ,\omega _{i2} )$ must be satisfied.
To achieve this condition for all $\omega$, firstly the two SPDC sources must be equal (indistinguishable):  $f_1 (\omega _{s} ,\omega _{i} ) =  f_2 (\omega _{s} ,\omega _{i} )=  f(\omega _{s} ,\omega _{i} )$, and secondly the two sources must be factorable (spectrally pure): $f (\omega _{s} ,\omega _{i} )= g_s (\omega _{s})  g_i (\omega _{i} )$ \cite{MosleyPhD}.

\begin{figure*}[tb]
\centering\includegraphics[width=15 cm]{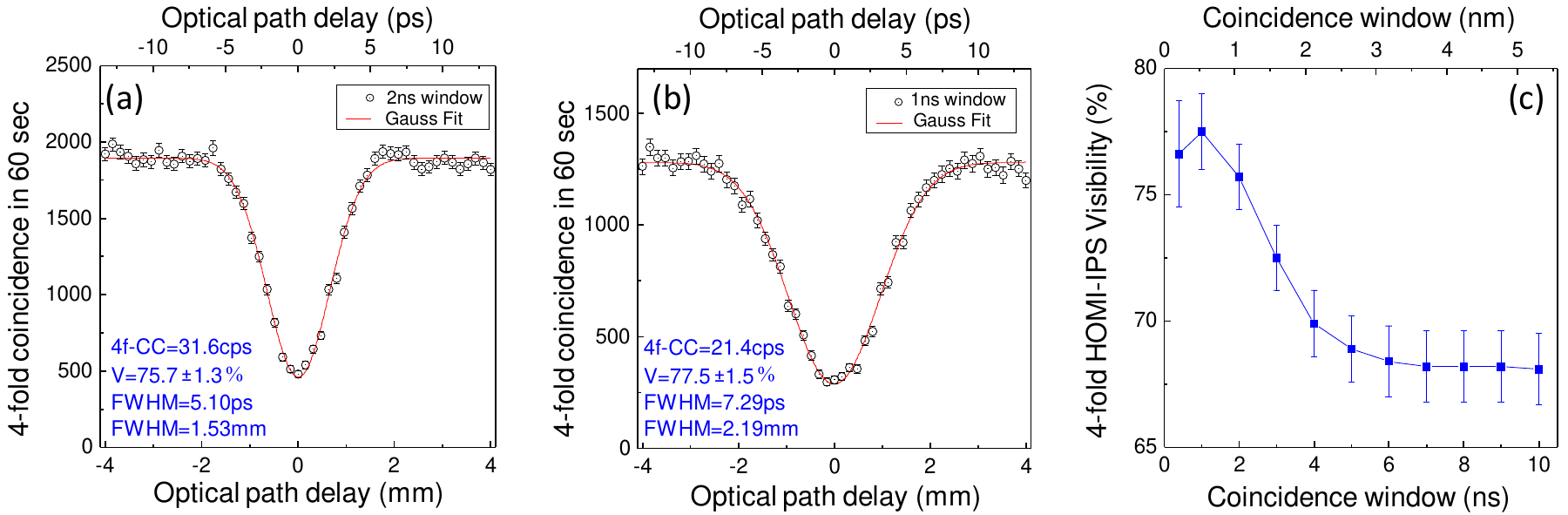}
\caption{ (a) Post-processed HOM dip with coincidence window of 2 ns (1 nm). (b) Post-processed HOM dip with coincidence window of 1 ns (0.5 nm). (c) Post-processed HOM dip visibility as a function of coincidence window.
  } \label{filter}
\end{figure*}

Besides the observation of the CSI, the technique in this work can also be applied to improve the visibility of HOMI-IPS by post-processing spectral filtering.
Here, we can improve the visibility from 68.9$\pm$1.3\% (in Fig. \ref{result1}(a)) to 75.7$\pm$1.3\% (in Fig. \ref{filter}(a)) and 77.5$\pm$1.5\% (in Fig. \ref{filter}(b)), by decreasing the coincidence window from 5 ns to 2 ns and 1 ns, corresponding to spectral widths from 2.66 nm to 1.06 nm and 0.53 nm, respectively.
The HOM interference visibilities with different coincidence window widths are shown in Fig. \ref{filter}(c).
Fig. \ref{filter}(c) clearly shows that the visibility increases as the width of the coincidence window decreases.
Thus, narrowing the coincidence window acts as a spectral filter.
When narrowing the coincidence window, the spectral purities of the sources are improved and the HOM visibility is improved.
Thus, one can utilize temporal filtering to enhance the visibility of the HOM-IPS.
However, at the cost of lower photon flux.

\section{Conclusion}
In summary, we have theoretically simulated and experimentally demonstrated the CSI from a four-fold HOM interference between two heralded single-photon sources.
We also demonstrated the CSI in a two-fold HOM interference between two thermal sources.
The interference pattern in the CSI is clearly observed and matches our theoretical predictions well.
This experiment can help enable the understanding of the HOMI-IPS in the spectral domain, and the technique can be utilized to improve the visibility of HOMI-IPS by temporal filtering.

\section*{Acknowledgments}
The authors are grateful to K. Yoshino and K. Wakui for assistance in experiment.
This work is supported by MEXT Grant-in-Aid for Young Scientists(B)15K17477, and the Program for Impulsing Paradigm Change through Disruptive Technologies (ImPACT).

\section*{Appendix: The theoretical model of the four-fold HOM interference}
This appendix briefly describe our theoretical model of the four-fold
HOM interference. The model includes losses in the setup,
the multi-pair emission (up to infinitely higher order), and
the frequency multimode structure of the pulsed photon-pair source.
Note that in  the theoretical model of heralded single-photon state  for Eq.(\ref{eq:P4}) and Eq.(\ref{eq:P2}), only the one-pair component in SPDC is considered.
While in the theoretical model of thermal state for Eq.(\ref{eq:A2}), both the one-pair  and two-pair components in SPDC are included.

An SPDC theory including the first two (losses and multi-pair
emission) was developed in \cite{Takeoka2015} where the quantum state emitted
from the SPDC source is described as a Gaussian continuous variable state,
more precisely two-mode squeezed vacuum, which is distributed in a single
frequency mode.
For pulsed SPDC sources, however, photon-pairs are emitted with more
complicated frequency distribution and thus has a multi-mode structure
in frequency. This is particularly important to simulate the experiments
including the interference between the photons from independent sources \cite{Walmsley2005, Lu2007a,Broome2013, Spring2013}.
Here we show how the frequency multimode structure can be incorporated
into the theory in \cite{Takeoka2015}
(complete description of the theory will appear elsewhere).

\begin{figure}[tbp]
\centering\includegraphics[width=8 cm]{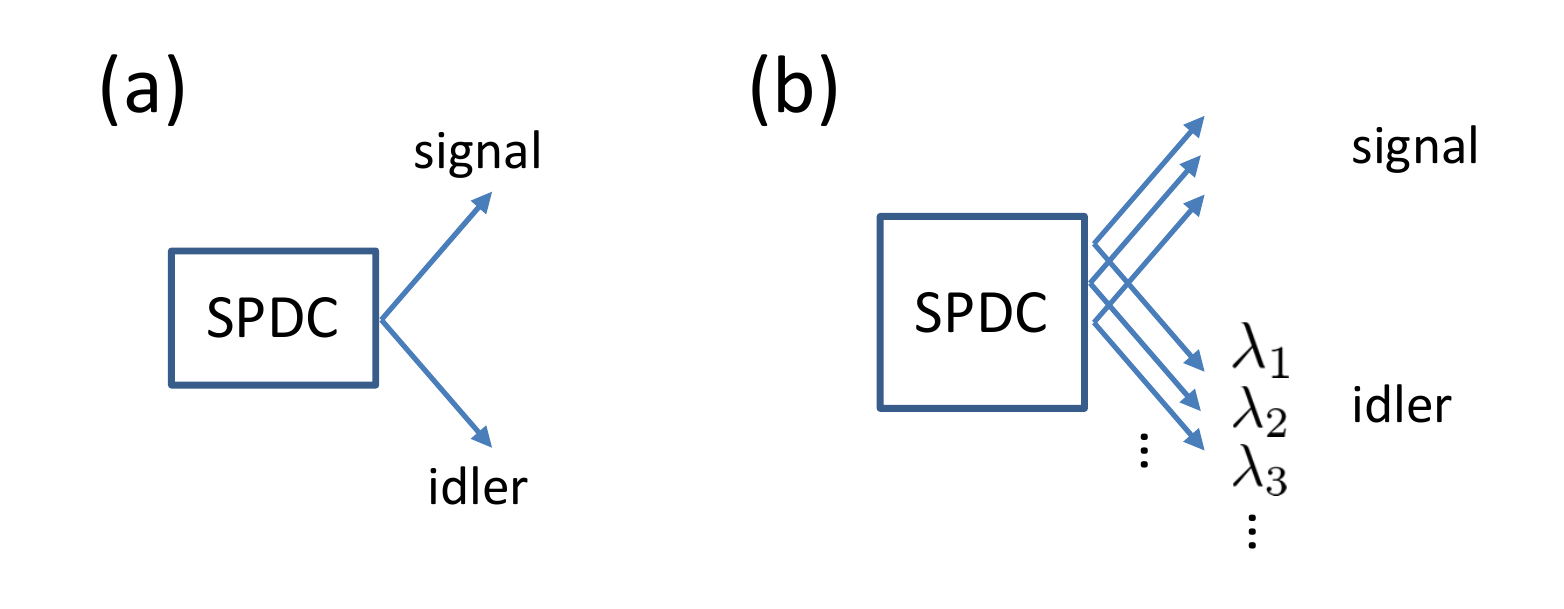}
\caption{ (a) single frequency mode SPDC source.
(b) pulsed (multiple frequency mode) SPDC source in Schmidt mode expansion.
  } \label{fig:Schmidt_mode}
\end{figure}

The quantum state emitted from a pulsed SPDC source is
described by \cite{Grice1997},
\begin{equation}
\label{eq:emitted_state}
|\Psi\rangle = \exp\left[ r \int d \omega_s \, \int d \omega_i
f(\omega_s, \omega_i) \hat{a}_s^\dagger (\omega_s)\hat{a}_i^\dagger (\omega_i)
- h.c. \right] |0\rangle ,
\end{equation}
where $r$ is the squeezing parameter (pump power),
$\hat{a}_s^\dagger (\omega_s)$ and $\hat{a}_i^\dagger (\omega_i)$
are creation operators for the signal and idler modes with frequencies
$\omega_s$ and $\omega_i$, respectively. $f(\omega_s, \omega_i)$ is
the joint spectral amplitude, which is a product of the pump distribution
and the phase-matching function of the nonlinear crystal
\cite{Grice1997,Law2000,Mosley2008a}.
We can assume $r$ to be real and positive without losing generality.
This joint spectral amplitude can be decomposed by the Schmidt decomposition
as \cite{Law2000,Mosley2008a}
\begin{equation}
f(\omega_s, \omega_i) = \sum_l \sqrt{\lambda_l} g_l(\omega_s) h_l(\omega_i) ,
\end{equation}
where $\lambda_l$ is the Schmidt eigenvalue for the $l$th Schmidt mode.
This decomposition allows us to describe the state
in Eq. (\ref{eq:emitted_state}) as
\begin{eqnarray}
\label{eq:Schmidt_decomposition}
|\Psi\rangle & = & \exp\left[ r \sum_l \sqrt{\lambda_l}
\hat{b}_l^\dagger \hat{c}_l^\dagger - h.c. \right] |0\rangle
\nonumber\\ & = &
\prod_l \exp\left[ r \sqrt{\lambda_l} \hat{b}_l^\dagger \hat{c}_l^\dagger
- h.c. \right] |0\rangle
\nonumber\\ & = &
|\Psi (r\sqrt{\lambda_1})\rangle|\Psi (r\sqrt{\lambda_2})\rangle \cdots ,
\end{eqnarray}
where
\begin{eqnarray}
\label{eq:b,c}
\hat{b}_l^\dagger & = & \int d\omega_s g_l (\omega_s)
\hat{a}_s^\dagger (\omega_s) ,
\\
\hat{c}_l^\dagger & = & \int d\omega_i h_l (\omega_i)
\hat{a}_i^\dagger (\omega_i) .
\end{eqnarray}
Note that $\hat{b}_l$ and $\hat{c}_l$ satisfy a standard bosonic
commutation relation
$[\hat{b}_l, \hat{b}_m^\dagger] = [\hat{c}_l, \hat{c}_m^\dagger]
= \delta_{lm}$.
In the second line of Eq.(\ref{eq:Schmidt_decomposition}),
decomposition of the exponential term is possible since
the Schmidt modes are orthonormal.
The last line of  Eq.(\ref{eq:Schmidt_decomposition}) represents
that in the Schmidt mode basis, the state is described by
a tensor product of two-mode squeezed vacuum
$|\Psi (r\sqrt{\lambda_l})\rangle$:
\begin{equation}
\label{eq:Scmidt_TMSV}
|\Psi (r\sqrt{\lambda_l})\rangle =
\frac{1}{\cosh r \sqrt{\lambda_l}}
\sum_n \left( \tanh r \sqrt{\lambda_l} \right)^n
|n\rangle_{B_l}|n\rangle_{C_l} ,
\end{equation}
with the effective squeezing parameter $r\sqrt{\lambda_l}$.
Therefore, to extend the single frequency mode based theory in \cite{Takeoka2015},
what need is simply consider a tensor product of multiple two-mode
squeezed vacua (see Fig. \ref{fig:Schmidt_mode}).

Distribution of the Schmidt eigenvalues are theoretically calculated
using the experimental parameters of our setup: $\sim$ 2 ps pump pulses at 792 nm, 30-mm-long   periodically poled $\mathrm{KTiOPO_4}$ (PPKTP) crystal.
Under this condition, it can be calculated that
$\{\lambda_1, \lambda_2, \dots \} = \{0.9, 0.025, 0.025, 0.01, 0.009, 0.004
\dots \}$  \cite{Jin2013OE, Branczyk2010, Grice2001}. In our numerical simulation, we include
the first five Schmidt modes and truncate the rest.

To fit the simulation with the experiment, we also need to connect
$r$ with the experimentally observable parameter, for example,
the photon-pair generation rate of the SPDC source. The latter corresponds
to the probability that the source emit non-zero photons:
\begin{equation}
\label{eq:pair_gen_rate}
p = 1 - \prod_l \left| \langle 00|\Psi(r\sqrt{\lambda_l})\rangle \right|^2 .
\end{equation}
Pluging Eq. (\ref{eq:Scmidt_TMSV}) into, Eq. (\ref{eq:pair_gen_rate}),
we obtain the relation
\begin{equation}
\label{eq:pair_gen_rate_vs_r}
p = 1 - \prod_l \left( \cosh r\sqrt{\lambda_l} \right)^{-2} ,
\end{equation}
which allows us to derive $r$ numerically from experimentally estimated $p$.
Note that the photon pair generation rate $p$ is related to
the mean photon number per pulse $\bar{n}$ via $p = \bar{n}/(\bar{n}+1)$.

The calculation of the four-fold HOM dip is then carried out by extending
the two-fold HOM dip analysis in \cite{Takeoka2015}.
Instead of a single TMSV in \cite{Takeoka2015}, we start from
the covariance matrix of two TMSVs and then after applying beam splitting
operation with mode matching factor $\xi$, we obtain the four-spatial-mode
covariance matrix $\gamma(r, \xi)$ where $r$ is the squeezing parameter
of the TMSVs (see \cite{Takeoka2015} for the details of the covariance matrix
approach). The HOM visibility is defined by
\begin{equation}
\label{eq:HOM_visibility}
V = \frac{P_{\rm mean} - P_{\rm min}}{P_{\rm mean}} ,
\end{equation}
where $P_{\rm mean}$ and $P_{\rm min}$ are the four-fold coincidence
count probabilities with and without optical path delay, respectively.
These coincidence counts are then given by
\begin{equation}
\label{eq:P_mean}
P_{\rm mean} = 1 + \sum_{m=1}^4 \, \sum_{C(4,m)}^{ _4 C_m}
\frac{(-2)^m}{\sqrt{ \prod_i
{\rm det} \left( \gamma_{C(4,m)}(r_i,0)+I \right)}} .
\end{equation}
$P_{\rm min}$ is also obtained by a similar expression with
$\gamma_{C(4,m)}(r_i,1)$ instead of $\gamma_{C(4,m)}(r_i,0)$.
$\gamma_{C(4,m)}(r,\xi)$ is the submatrix of $\gamma(r,\xi)$ taking
modes $C(4,m)$. $C(4,m)$ is the $m$-detector combination from the four
detectors (see Sec. 3.3 of \cite{Takeoka2015}).
$r_i=r\sqrt{\lambda_i}$ is the effective squeezing parameter
for the $i$th Schmidt mode where we include the first five modes.
The HOM visibility is then obtained by plugging the above $P_{\rm mean}$ and
$P_{\rm min}$ into Eq. (\ref{eq:HOM_visibility}).

\end{document}